\documentclass[10pt]{IEEEtran}

\usepackage{amsmath}
\usepackage{amssymb}

\usepackage{cite}

\def\C{{\mathbb C}}
\def\F{{\mathbb F}}
\def\dist{\mathop{\rm dist}}
\def\wgt{\mathop{\rm wgt}}

\def\ket#1{|#1\rangle}

\newtheorem{definition}{Definition}
\newtheorem{proposition}[definition]{Proposition}

\newtheorem{theorem}[definition]{Theorem}

\newtheorem{example}[definition]{Example}

\def\squareforqed{\hbox{\rlap{$\sqcap$}$\sqcup$}}
\def\qed{\ifmmode\squareforqed\else{\unskip\nobreak\hfil
\penalty50\hskip1em\null\nobreak\hfil\squareforqed
\parfillskip=0pt\finalhyphendemerits=0\endgraf}\fi}

\begin{document}

\title{Generalized Concatenation for Quantum Codes}

\author{\IEEEauthorblockN{\large Markus Grassl\IEEEauthorrefmark{1}\IEEEauthorrefmark{2},
Peter W. Shor\IEEEauthorrefmark{3}, and
Bei Zeng\IEEEauthorrefmark{4}
}
\bigskip

\IEEEauthorblockA{\IEEEauthorrefmark{1}Institute for Quantum Optics and Quantum Information,
Austrian Academy of Sciences,\\
Technikerstra{\ss}e 21a, 6020 Innsbruck, Austria\\[0.25ex]
\IEEEauthorrefmark{2}Centre for Quantum Technologies,
National University of Singapore,\\
3 Science Drive 2, Singapore 117543,
Singapore, Email: Markus.Grassl@nus.edu.sg\medskip
}

\IEEEauthorblockA{\IEEEauthorrefmark{3}Department of Mathematics,
Massachusetts Institute of Technology,\\
Cambridge, MA 02139, USA, Email: shor@math.mit.edu\medskip}

\IEEEauthorblockA{\IEEEauthorrefmark{4}Department of Physics,
Massachusetts Institute of Technology,\\
Cambridge, MA 02139, USA, Email: zengbei@mit.edu}
}

\maketitle

\pagenumbering{arabic}

\begin{abstract}
We show how good quantum error-correcting codes can be constructed
using generalized concatenation.  The inner codes are quantum codes,
the outer codes can be linear or nonlinear classical codes.  Many new
good codes are found, including  both stabilizer codes as well as so-called
nonadditive codes.
\end{abstract}

\begin{IEEEkeywords}
Generalized concatenated codes,
quantum error correction,
stabilizer codes,
nonadditive codes  
\end{IEEEkeywords}

\section{Introduction}
The idea  of concatenated codes,  originally described by Forney  in a
seminal  book   in  1966  \cite{For66},  was   introduced  to  quantum
computation   three  decades   later  \cite{AhBO97,KLZ96,KLZ98,Got97}.
These concatenated quantum codes play a central role in fault tolerant
quantum computation  (FTQC) as  well as in  the study  of constructing
good degenerate quantum codes.

Blokh and Zyablov \cite{BlZy74}, followed by Zinoviev \cite{Zin76}
introduced the concept of generalized concatenated codes.  These codes
improve the parameters of conventional concatenated codes for short
block lengths \cite{Zin76} as well as their asymptotic performance
\cite{BlZy82}. Many good classical codes, linear and nonlinear, can be
constructed using this method.

In \cite{GSSSZ09} we, together with Smith and Smolin, have introduced
generalized concatenated quantum codes (GCQC).  It is shown that GCQC
in its simplest form, i.\,e., two level concatenation, is already a
powerful tool to produce good nonadditive quantum codes which
outperform any stabilizer codes.

This paper focuses on the multilevel concatenation for quantum codes.
We use the framework of stabilizer codes and the generalization to
codeword stabilized (CWS) codes \cite{CSSZ09,CZC08} and union
stabilizer codes \cite{GrRo08:ITW,GrRo08:ISIT}.  This allows to use
classical codes as outer codes.  We further extend our multilevel
concatenation technique to the case of different inner codes, which
allows us to construct codes of various lengths.

\section{Background and Notations}
A general quantum error-correcting code (QECC), denoted by
$C=((n,K,d))_q$, is a $K$-dimension subspace of the Hilbert space
$\mathcal{H}_q^{\otimes n}$ of dimension $q^n$ that is the tensor
product of $n$ complex Hilbert spaces $\mathcal{H}_q=\C^q$ of
dimension $q$. Here we restrict $q=p^m$ to be a prime power.  A QECC
with minimum distance $d$ allows to correct arbitrary errors that
affect at most $(d-1)/2$ of the $n$ subsystems.

Most of the known QECCs are so-called stabilizer codes introduced
independently by Gottesman \cite{Got96} and Calderbank et
al. \cite{CRSS98}.  The code is defined as the joint eigenspace of a
set of commuting operators \cite{Got96}. Equivalently, the code can be
described by a classical additive code $\mathcal{C}$ over $GF(q^2)$
that is self-orthogonal with respect to a symplectic inner product
\cite{CRSS98,AsKn01}.  Denoting the symplectic dual code by
$\mathcal{C}^*$, the minimum distance of the quantum code is given by
\begin{alignat*}{3}
d=\min\{\wgt(c)\colon c\in \mathcal{C}^*\setminus \mathcal{C}\}\ge d_{\min}(\mathcal{C}^*).
\end{alignat*}
If $d=d_{\min}(\mathcal{C}^*)$, the quantum code is called pure or
nondegenerate.  The corresponding stabilizer (or additive) code is
denoted by $C=[[n,k,d]]_q$ and has dimension $K=q^k$.

The first nonadditive code $((5,6,2))_2$ which has a higher dimension
than any stabilizer code of the same length correcting one erasure can
be explained as the union of six locally transformed copies of the
stabilizer code $[[5,0,3]]_2$ (see \cite{RHSS97,GrBe97}).  A
one-dimensional stabilizer code $[[n,0,d]]$ can also be described by a
graph with $n$ vertices \cite{ScWe02}. The corresponding quantum
states are referred to as graph states.  Combining locally equivalent
graph states, the first one-error-correcting nonadditive quantum code
$((9,12,3))_2$ with higher dimension than any stabilizer code has
been found \cite{YCL08}.  The theoretical ground for these codeword
stabilized (CWS) quantum codes has been laid in \cite{CSSZ09,CZC08}.

In \cite{GrRo08:ITW,GrRo08:ISIT}, the framework of union stabilizer
codes has been introduced.  Starting with a stabilizer code
$C_0=[[n,k,d_0]]_q$, a union stabilizer code is given by
\begin{alignat*}{3}
C=\bigoplus_{t\in T_0} t C_0,
\end{alignat*}
where $T_0=\{t_1,\ldots,t_K\}$ is a set of tensor products of
(generalized) Pauli matrices such that the spaces $t_i C_0$ are
mutually orthogonal. Then the dimension of the union stabilizer code
$C$ is $K q^k$, and we will use the notation ${\cal C}=((n,Kq^k,d))_q$.  
Similar to stabilizer codes, a union stabilizer code can be described
in terms of classical codes.  Given the symplectic dual
$\mathcal{C}_0^*$ of the additive code $\mathcal{C}_0$ associated to
the stabilizer code $C_0$, the union normalizer code is the union of
cosets of $\mathcal{C}_0^*$ given by
\begin{equation}\label{eq:unionnormalizer}
\mathcal{C}^*=\bigcup_{t\in {\mathcal{T}}_0} \mathcal{C}_0^*+t=\{c+t_j\colon c \in \mathcal{C}_0^*,\, j=1,\ldots,K\}.
\end{equation}
Here $\mathcal{T}_0$ is the set of vectors $t_i\in\F^n_{q^2}$
corresponding to the generalized Pauli matrices $t_i\in T_0$.
\begin{proposition}[cf. \cite{GrRo08:ISIT}]\label{prop:unionmindist}
The minimum distance of a union stabilizer code with union normalizer
code $\mathcal{C}^*$ is given by
\begin{alignat*}{5}
d&=\min\{\wgt(v)\colon v \in(\mathcal{C}^*-\mathcal{C}^*)\setminus\widetilde{\mathcal{C}}_0\}\kern-50pt\\
 &\ge d_{\min}(\mathcal{C}^*)\\
&=\min\{\dist(c+t_i,c'+t_{i'})\colon t_i,t_{i'}\in{\cal T}_0,\,&& c,c'\in \mathcal{C}^*_0\\
&&& c+t_i\ne c'+t_{i'}\},
\end{alignat*}
where $\mathcal{C}^*-\mathcal{C}^*:=\{a-b\colon a,b \in
\mathcal{C}^*\}$ denotes the set of all differences of vectors in
$\mathcal{C}^*$, and $\widetilde{\mathcal{C}}_0\le \mathcal{C}_0$ is
the symplectic dual of the additive closure of the (in general
nonadditive) union normalizer code $\mathcal{C}^*$.
\end{proposition}

Hence in order to construct a union stabilizer code with distance $d$,
it suffices to find a large classical code $\mathcal{C}^*$ with
minimum distance $d$ that can be decomposed into cosets of an additive
code $\mathcal{C}^*_0$ that contains its symplectic dual.  Two
extremal cases are stabilizer codes where only one coset is used, and
CWS codes for which $\mathcal{C}^*_0=\mathcal{C}_0$ is a symplectic
self-dual code.

\section{Generalized Concatenation}
The basic idea of generalized concatenated quantum codes
\cite{GSSSZ09} uses just two levels of concatenation.  Here we first
present multilevel concatenation for quantum codes. Then we discuss a
special case that can be described by classical codes only.

\subsection{Multilevel Concatenation for Quantum Codes}
The inner quantum code ${B}^{(0)}=((n,q_1q_2\cdots q_r,d_1))_q$ is
first partitioned into $q_1$ mutually orthogonal subcodes
${B}^{(1)}_{i_1}$ ($0\leq i_1\leq q_1-1$), where each $B^{(1)}_{i_1}$
is an $((n,q_2\cdots q_r,d_2))_q$ code. Then each ${B}^{(1)}_{i_1}$ is
partitioned into $q_2$ mutually orthogonal subcodes
${B}^{(2)}_{i_1i_2}$ ($0\leq i_2\leq q_2-1$), where
${B}^{(2)}_{i_1i_2}$ has parameters $((n,q_3\cdots q_r,d_3))_q$, and
so on.  Finally, each ${B}^{(r-2)}_{i_1i_2\ldots i_{r-2}}$ is
partitioned into $q_{r-1}$ mutually orthogonal subcodes
${B}^{(r-1)}_{i_1i_2\ldots i_{r-1}}=((n,q_r,d_r))_q$ for $0\leq i_{r-1} \leq
q_{r-1}-1$.  Thus
\begin{alignat}{7}
{B}^{(0)}=\bigoplus_{i_1=0}^{q_1-1}{B}^{(1)}_{i_1},&\quad &
{B}^{(1)}_{i_1}=\bigoplus_{i_2=0}^{q_2-1}{B}^{(2)}_{i_1i_2},&\quad &
\ldots,\label{eq:inner_dec}
\end{alignat}
and $d_1\le d_2\le\ldots\le d_r$.  A typical basis vector of
${B}^{(0)}$ will be denoted by $\ket{\varphi_{i_1i_2\ldots i_r}}$
$(0\leq i_1\leq q_1-1,\ldots,0\leq i_r\leq q_r -1)$, with subscripts
chosen such that $\ket{\varphi_{i_1i_2\ldots i_r}}$ is a basis vector
of all ${B}_{i_1}^{(1)},
{B}^{(2)}_{i_1i_2},\ldots,{B}^{(r-1)}_{i_1i_2\ldots i_{r-1}}$.

In addition, we take as outer codes a collection of $r$ quantum codes
${A}_1,\ldots,{A}_r$, where ${A}_j$ is an $((N,M_j,\delta_j))_{q_j}$
code over the Hilbert space $\mathcal{H}_{q_j}^{\otimes N}$.  Denote
the standard basis of each $\mathcal{H}_{q_j}^{\otimes N}$ by
\[
\{\ket{i_1^{(j)}}\otimes\ldots\otimes\ket{i_N^{(j)}}\colon 0\le i^{(j)}_\nu\le q_j-1,1\le\nu\le N\}
\]
(where $j$ runs from $1$ to $r$), and the bases of the codes $A_j$ are
denoted by $\{\ket{\phi^{(j)}_{l_j}}\colon 0\le l_j \le M_j-1\}$.
Expanding the basis vectors of $A_j$ with respect to the standard
basis of $\mathcal{H}_j^{\otimes N}$ we obtain
\begin{alignat}{5}
\ket{\phi^{(j)}_{l_j}}=\!\!
\sum_{i^{(j)}_1 i^{(j)}_2\ldots  i^{(j)}_N}\alpha^{(j)}_{l_j,i^{(j)}_1 i^{(j)}_2\ldots i^{(j)}_N}\!\!
  \ket{i^{(j)}_1}\otimes\ket{i^{(j)}_2}\otimes\ldots\otimes\ket{i^{(j)}_N}.
\label{eq:basis_outercode}
\end{alignat}
The basis vectors of the tensor product of all outer codes are given
by
\begin{alignat*}{5}
\ket{\phi^{(1)}_{l_1}}\otimes\ket{\phi^{(2)}_{l_2}}\otimes\ldots\otimes\ket{\phi^{(r)}_{l_r}},
\end{alignat*}
where $l_j$ runs from $0$ to $M_j-1$. Expanding these basis vectors
with respect to the standard bases we obtain
\begin{alignat}{5}
\rlap{\kern-4mm$\ket{\phi^{(1)}_{l_1}}\otimes\ket{\phi^{(2)}_{l_2}}\otimes\dots\otimes\ket{\phi^{(r)}_{l_r}}=$}\nonumber\\
&\left(\sum_{i^{(1)}_1 i^{(1)}_2\ldots  i^{(1)}_N}\alpha^{(1)}_{l_1,i^{(1)}_1 i^{(1)}_2\ldots i^{(1)}_N}
  \ket{i^{(1)}_1}\otimes\ket{i^{(1)}_2}\otimes\ldots\otimes\ket{i^{(1)}_N}\right)\nonumber\\
\otimes&\left(\sum_{i^{(2)}_1 i^{(2)}_2\ldots  i^{(2)}_N}\alpha^{(2)}_{l_2,i^{(2)}_1 i^{(2)}_2\ldots i^{(2)}_N}
  \ket{i^{(2)}_1}\otimes\ket{i^{(2)}_2}\otimes\ldots\otimes\ket{i^{(2)}_N}\right)\nonumber\\[1ex]
\ldots\nonumber\\[1ex]
\otimes&\left(\sum_{i^{(r)}_1 i^{(r)}_2\ldots  i^{(r)}_N}\alpha^{(r)}_{l_r,i^{(r)}_1 i^{(r)}_2\ldots i^{(r)}_N}
  \ket{i^{(r)}_1}\otimes\ket{i^{(r)}_2}\otimes\ldots\otimes\ket{i^{(r)}_N}\right).\label{outerbasis}
\end{alignat}
The basis of the resulting generalized concatenated quantum code $Q$
is given by replacing the basis vectors in Eq. (\ref{outerbasis})
using the mapping
\begin{alignat*}{5}
\ket{i^{(1)}_\nu}\otimes\ket{i^{(2)}_\nu}\otimes\ldots\otimes\ket{i^{(r)}_\nu}
\mapsto\ket{\varphi_{i^{(1)}_\nu i^{(2)}_\nu \ldots i^{(r)}_\nu}}
\end{alignat*}
for $1\le\nu\le N$.  Hence the basis of $Q$ is given by
\begin{alignat*}{5}
\ket{\psi_{l_1l_2\ldots l_r}}&=\sum&&
  \alpha^{(1)}_{l_1,i^{(1)}_1 i^{(1)}_2\ldots i^{(1)}_N}\cdots
  \alpha^{(r)}_{l_r,i^{(r)}_1 i^{(r)}_2\ldots i^{(r)}_N}\\
&&&\qquad\ket{\varphi_{i^{(1)}_1 i^{(2)}_1 \ldots i^{(r)}_1}}
\otimes\ldots\otimes
\ket{\varphi_{i^{(1)}_N i^{(2)}_N \ldots i^{(r)}_N}}.
\end{alignat*}
So $Q$ is a quantum code in the Hilbert space
$\mathcal{H}_{q}^{\otimes Nn}$ of dimension $M=M_1M_2\cdots M_r$.  As
already mentioned, the construction given in \cite{GSSSZ09} is a
two-level construction with $r=2$, while the concatenation of quantum
codes used in the context of fault tolerant quantum computation
(cf. \cite{AhBO97,KLZ96,KLZ98,Got97}) is a one-level construction,
i.\,e. $r=1$.

\subsection{Classical Outer Codes}
From now on we restrict ourselves in constructing union
stabilizer codes.  For simplicity we consider only nondegenerate codes
here.

We take the inner code $B^{(0)}$ to be an $((n,K q^k,d_1))_q$
nondegenerate union stabilizer code, given by a classical symplectic
self-orthogonal additive code $\mathcal{C}_0\subset
\mathcal{C}^*_0=(n,q^{n+k},d_r)_{q^2}$ and a set $\mathcal{T}^{(0)}$
of $K=q_1 q_2\cdots q_{r-1}$ coset representatives.  The corresponding
classical union normalizer code is
\[
\mathcal{C}^*=\mathcal{B}^{*(0)}=\bigcup_{t\in\mathcal{T}^{(0)}}
\mathcal{C}^*_0+t.
\]
The decomposition (\ref{eq:inner_dec}) of the inner quantum code
$B^{(0)}$ into mutually orthogonal union stabilizer codes is based on
the decomposition of the union normalizer code $\mathcal{B}^{*(0)}$
that is obtained by partitioning the coset representatives
\begin{alignat*}{5}
\mathcal{T}^{(0)}=\bigcup_{i_1=0}^{q_1-1}\mathcal{T}^{(1)}_{i_1},&\quad&
\mathcal{T}^{(1)}_{i_1}=\bigcup_{i_2=0}^{q_2-1}\mathcal{T}^{(2)}_{i_1i_2},&\quad&\ldots
\end{alignat*}
This defines union normalizer codes $\mathcal{B}^{*(j)}$ given by
\[
\mathcal{B}_{i_1i_2\ldots i_{j-1}}^{*(j)}=\bigcup_{t\in\mathcal{T}^{(j)}_{i_1i_2\ldots i_{j-1}}} \mathcal{C}^*_0+t.
\]
The coset representatives in $\mathcal{T}^{(0)}$ will be denoted by
$t_{i_1i_2\ldots i_{r-1}}$ with $0\leq i_1\leq q_1-1$, \ldots, $0\leq i_{r-1}\leq
q_{r-1}-1$.  The indices are chosen such that $t_{i_1i_2\ldots i_{r-1}}$
belongs to all
$\mathcal{T}_{i_1}^{(1)},\mathcal{T}^{(2)}_{i_1i_2},\ldots,\mathcal{T}^{(r-2)}_{i_1i_2\ldots
  i_{r-2}}$. 

Here $\mathcal{B}^{*(0)}$ is a classical code over $GF(q^2)$ with
parameters $(n,q_1q_2\cdots q_{r-1}q^{n+k},d_1)_{q^2}$ that is the
union of $q_1$ disjoint codes $\mathcal{B}^{*(1)}_{i_1}=(n,q_2\cdots
q_{r-1}q^{n+k},d_2)_{q^2}$, and so on.  Finally, each
$\mathcal{B}^{*(r-2)}_{i_1i_2\ldots i_{r-2}}$ is the union of
$q_{r-1}$ disjoint codes $\mathcal{B}^{*(r-1)}_{i_1i_2\ldots
  i_{r-1}}=(n,q^{n+k},d_r)_{q^2}$, each of which is a single coset of
the additive code $\mathcal{C}^*_0$.

In total we use $r$ classical outer codes. For the first $r-1$ outer
codes we take $\mathcal{A}_i=(N,M_i,\delta_i)_{q_i}$, a classical code
over an alphabet of size $q_i$ with length $N$, size $M_i$, and
distance $\delta_i$. The code $\mathcal{A}_r$ is a trivial code
$\mathcal{A}_r=[N,N,1]_{q_r}$ where $q_r=|\mathcal{C}^*_0|=q^{n+k}$.

Next we show how to construct the classical generalized concatenated
code using the inner code $\mathcal{B}^{*(0)}$ and the outer codes
$\mathcal{A}_1,\ldots,\mathcal{A}_r$.  What follows is an adaption of
\cite[Ch.~18, \S8.2]{MS77}.  The trivial classical code
$\mathcal{A}_r=[N,N,1]_{q_r}$ on level $r$ is concatenated with the
additive normalizer code $\mathcal{C}^*_0$, resulting in the additive
code $(\mathcal{C}^*_0)^N$ which contains its symplectic dual
$\mathcal{C}_0^N$.  Note that this corresponds to concatenating a
trivial quantum code $A_r=[[N,N,1]]_{q^k}$ with the stabilizer code
$C_0$.  As a technicality we note that the alphabet size of the
trivial classical outer code $\mathcal{A}_r$ is $q^{n+k}$, while the
trivial outer quantum code $A_r$ is over quantum systems of dimension
$q^k$.

The first $r-1$ outer codes are used to define a set of coset
representatives. For this, form an $N\times (r-1)$ array
\[
\left[ 
\begin{array}{cccc}
a_1^{(1)} & a_1^{(2)} & \cdots & a_1^{(r-1)} \\
a_2^{(1)} & a_2^{(2)} & \cdots & a_2^{(r-1)} \\
\vdots  & \vdots   & \ddots & \vdots   \\
a_N^{(1)} & a_N^{(2)} & \cdots & a_N^{(r-1)}
\end{array}
\right],
\] 
where the first column is a codeword of $\mathcal{A}_1$, the second is
in $\mathcal{A}_2$, etc.  Then replace each row $a_j^{(1)}, a_j^{(2)},
\ldots, a_j^{(r-1)}$ by the coset representative $t_{a_j^{(1)},
  a_j^{(2)}, \ldots, a_j^{(r-1)}}=T_j$.  (For this, label the elements
of the alphabet of size $q_i$ by the numbers $0,1,\ldots, q_i-1$ in
some arbitrary, but fixed way.)  The resulting $N\times n$ arrays
$T=(T_1,\ldots,T_N)$ (considered as vectors of length $Nn$) form the
new set of coset representatives of the generalized concatenated code
\begin{equation}\label{eq:classical_gc}
\mathcal{C}^*_{\text{gc}}
=\bigcup_{(T_1,\ldots,T_N)} (\mathcal{C}^*_0\times\ldots\times\mathcal{C}^*_0)+(T_1,\ldots,T_N).
\end{equation}
Clearly, this code $\mathcal{C}^*_{\text{gc}}$ has the form of a union
normalizer code as specified in (\ref{eq:unionnormalizer}).  Hence
$\mathcal{C}^*_{\text{gc}}$ defines a QECC.  The properties of this code are
given by the following theorem.
\begin{theorem}
The minimum distance of the union normalizer code
\[
C_{\text{gc}}=((nN,M_1 M_2\cdots M_{r-1}q^{kN},d))_q,
\]
corresponding to $\mathcal{C}^*_{\text{gc}}$ given in (\ref{eq:classical_gc})
is
\[
d\geq \min\{\delta_1d_1,\ldots,\delta_{r-1}d_{r-1},d_r\}.
\]
\end{theorem}
\begin{IEEEproof} 
Let $c$ and $\tilde{c}$ be two distinct codewords of
$\mathcal{C}^*_{\text{gc}}$.  If they belong to the same coset, then
$c-\tilde{c}\in (\mathcal{C}^*_0)^N$.  Hence their distance is at
least $d_r$.  Now assume that $c$ and $\tilde{c}$ lie in different
cosets given by the arrays $(a^{(i)}_j)$ and $(\tilde{a}^{(i)}_j)$.
If the arrays differ in the $\nu^{\text{th}}$ column then they differ
in at least $\delta_{\nu}$ places in the $\nu^{\text{th}}$ column.  By
definition $t_{i_1i_2\ldots i_{\nu-1}\alpha\ldots}$ and
$t_{i_1i_2\ldots i_{\nu-1}\beta\ldots}$ (with $\alpha\neq\beta$) both
belong to $\mathcal{T}^{(\nu-1)}_{i_1i_2\ldots i_{\nu-1}}$.  Therefore
the corresponding codewords of $\mathcal{B}^{*(\nu-1)}_{i_1i_2\ldots
  i_{\nu-1}}$ differ in at least $d_{\nu}$ places.  Hence $c$ and
$\tilde{c}$ differ in at least $\delta_{\nu}d_{\nu}$ places.
\end{IEEEproof}

\subsection{Additivity Properties}
We know that if $\mathcal{C}^*_{\text{gc}}$ is an additive code, then the
corresponding quantum code $C_{\text{gc}}$ is a stabilizer code.  So the
question is when does generalized concatenation yield an additive
code. The following is an adaption of a result from \cite{Dum98}.
\begin{proposition}
Given additive, i.\,e., $\F_p$-linear, outer codes
$\mathcal{A}_1,\ldots,\mathcal{A}_{r-1}$ and an additive inner code
$\mathcal{B}$, the resulting generalized concatenated code is
additive if the mapping
\begin{equation}\label{eq:additive_mapping}
(a_i^{(1)}, a_i^{(2)}, \ldots, a_i^{(r-1)})\mapsto t_{a_i^{(1)}, a_i^{(2)}, \ldots, a_i^{(r-1)}} 
\end{equation}
is $\F_p$-linear.
\end{proposition}
Hence we can construct stabilizer codes from a sequence of nested
stabilizer codes yielding a decomposition of the inner code and
classical linear outer codes.
\begin{theorem}\label{theorem:gc_stabilizer}
Let 
\begin{alignat*}{5}
B^{(0)}=[[n,k_0,d_1]]_q &\supset B^{(1)}=[[n,k_1,d_2]]_q \supset\ldots\\
\ldots&\supset B^{(r-1)}=[[n,k_{r-1},d_r]]_q
\end{alignat*}
be a sequence of nested nondegenerate stabilizer codes.  This defines
a decomposition of the inner code $B^{(0)}$.  Using $r-1$ additive
outer codes $\mathcal{A}_i=(N,M_i,\delta_i)_{q_i}$ where
$q_i=q^{k_{i-1}-k_i}$ together with the trivial code
$\mathcal{A}_r=[N,N,1]_{q_r}$ where $q_r=q^{n+k_{r-1}}$, by
generalized concatenation we obtain a stabilizer code with parameters
$[[nN,K,d]]_q$ where
\[
d\ge\min\{ \delta_1 d_1,\delta_2 d_2,\ldots,\delta_{r-1} d_{r-1},d_r\}
\]
and
\[
K=k_r^N\log_q(M_1 M_2\cdots M_{r-1}).
\] 
\end{theorem}
Examples for this theorem are given in the next section.

\section{Examples}
\subsection{Stabilizer Codes}
\begin{example}
Consider the following sequence of nested stabilizer codes:
\[
B^{(0)}=[[6,6,1]]_2 \supset B^{(1)}=[[6,4,2]]_2 \supset B^{(2)}=[[6,0,4]]_2.
\]
The largest code $B^{(0)}$ can be decomposed into four mutually
orthogonal subspaces, each of which is a code $[[6,4,2]]_2$. Then each
of these codes $B^{(1)}$ is decomposed into 16 one-dimensional spaces
$[[6,0,4]]_2$.  Hence we need nontrivial outer codes with alphabet
sizes $4$ and $16$, which we chose to be
\[
\mathcal{A}_1=[6,3,4]_4\quad\text{and}\quad\mathcal{A}_2=[6,5,2]_{16},
\]
together with $\mathcal{A}_3=[6,6,1]_{2^6}$.  The dimension of the
resulting code is
$|\mathcal{A}_1|\times|\mathcal{A}_2|=4^316^5=2^62^{20}=2^{26}$, and
the minimum distance is at least $\min\{4\times 1,2\times 2,4\}=4$.
Taking an additive map (\ref{eq:additive_mapping}), we obtain a
stabilizer code.  As all inner codes are $GF(4)$-linear, we can even
chose the mapping (\ref{eq:additive_mapping}) to be $GF(4)$-linear,
resulting in a $GF(4)$-linear code $[[36,26,4]]_2$.  This code
improves the lower bound on the minimum distance of a stabilizer code
$[[36,26,d]]_2$ given in \cite{Grassl:codetables}.
\end{example}

Our construction allows to adopt most of the known variations of
generalized concatenation for classical codes.  In \cite{DGS95} a
modified generalized concatenation has been introduced which uses
outer code $\mathcal{A}_i$ of different lengths $n_i$ as well as
different inner codes $B_j^{(0)}$.
\begin{example}
Using the stabilizer code $B^{(1)}=[[21,15,3]]_2$, we can decompose
the full space $B^{(0)}=[[21,21,1]]_2$ into $64$ mutually orthogonal
codes $[[21,15,3]]_2$.  In order to construct a generalized
concatenated quantum code of distance three, we need a classical
distance-three code over an alphabet of size 64, e.\,g., the classical
MDS code $\mathcal{A}_1=[65,63,3]_{2^6}$, as well as the trivial code
$\mathcal{A}_2=[65,65,1]_{2^{21+15}}$.  Then by generalized
concatenation one obtains a perfect quantum code $[[1365,1353,3]]_2$.
Instead of taking $65$ copies of the inner code of length $21$, we can
use any combination of inner codes $B_j^{(1)}=[[n_j,n_j-6,3]]_2$ with
$n_j\in\{7,\ldots,17,21\}$.  Note that now the trivial outer code
$\mathcal{A}_2$ has to be modified in such a way that by concatenation
we get the normalizer code of the direct product of the various inner
codes $B_j^{(1)}$.  Overall we obtain quantum codes with parameters
$[[n,n-12,3]]_2$ for $n=455,\ldots,1361$ and $n=1365$.

Note that for quantum codes, the existence of a code $[[n,k,d]]_q$
does not necessarily imply the existence of a shortened code
$[[n-s,k-s,d]]_q$.  In general, one would have to analyze the weight
structure of an auxiliary code, the so-called puncture code,
introduced in \cite{Rai99}.  Varying the length of the inner quantum
codes, we can directly construct shorter codes.
\end{example}

\subsection{Nonadditive Codes}
In our construction, we can also use classical nonlinear codes as
outer codes.  Good nonlinear codes can be obtained as subcodes of a
linear code over a larger alphabet (or one of its cosets) by taking
only those codewords whose symbols are taken from a subset of the
alphabet.  The following result can be found in \cite[Lemma
  3.1]{Dum98}):
\begin{proposition}\label{prop:subalphabetcode}
If there exists an $(n,K,d)_q$ code, then for any
$s<q$, there exists an $(n',K',d)_s$ code with size at least
$K(s/q)^n$.
\end{proposition}

\begin{example}[cf. \cite{GSSSZ09}]
We start with the sequence of inner codes
\[
B^{(0)}=[[5,5,1]]_2 \supset B^{(1)}=[[5,1,3]]_2.
\]
For the nontrivial outer code we take a code over an alphabet of size
$2^{5-1}=16$ and distance three.  From the linear MDS code
$[18,16,3]_{17}$ over $GF(17)$ we can derive a nonlinear code
$\mathcal{A}_1=(18,\lceil \frac{16^{18}}{17^2}\rceil,3)_{16}$ over
$GF(16)$ using Proposition \ref{prop:subalphabetcode}.  The resulting
generalized concatenated quantum code has parameters
$((90,2^{81.825},3))_2$, while the best stabilizer code has parameters
$[[90,81,3]]_2$.
\end{example}
In the final example, we use three levels of concatenation and a
nonlinear classical outer code.
\begin{example}
Decompose the code $B^{(0)}=[[8,8,1]]_2$ using the sequence of
nested stabilizer codes
\[
B^{(0)}=[[8,8,1]]_2\supset B^{(1)}=[[8,6,2]]_2
\supset B^{(2)}=[[8,3,3]]_2.
\]
As outer codes we need a code with alphabet size $2^{8-6}=4$ and
distance three, a code with alphabet size $2^{6-3}=8$ and distance
two, as well as a trivial code.  We take the nonlinear code
$\mathcal{A}_1=(6,\lceil 4^6/5^2\rceil ,3)_{4}$ derived from the
linear MDS code $[6,4,3]_5$ over $GF(5)$, the linear code
$\mathcal{A}_2=[6,5,2]_{8}$ over $GF(8)$, and the linear code
$\mathcal{A}_3=[6,6,1]_{2^{8+3}}$.  The dimension of the generalized
concatenated quantum code is
$|\mathcal{A}_1|\times|\mathcal{A}_2|\times \dim(B^{(2)})^6=164\times
8^5\times 2^{3\times 6}$. Hence we get a nonadditive code
$((48,2^{40.356},3))_2$, which has a higher dimension than the best
possible additive code $[[48,40,3]]_2$.
\end{example}

\section{Decoding}
One of the advantages of concatenated codes as well as generalized
concatenated codes is that decoding can be based on decoding
algorithms for the constituent codes \cite{Dum98,For66}.  For quantum
codes, however, it is not possible to directly measure the ``code
symbols''. Instead, decoding is based on measuring an error syndrome.

For stabilizer codes, the error syndrome is obtained by measuring the
eigenvalues of generators of the stabilizer group.  The error syndrome
can be defined in such a way that it corresponds to the error syndrome
of the underlying classical code, and hence a classical decoding
algorithm can be used.

For generalized concatenated quantum codes derived from a sequence of
nested stabilizer codes as in Theorem~\ref{theorem:gc_stabilizer}, the
corresponding stabilizer groups are nested as well, with the
stabilizer of the smallest code $B^{(r)}$ being the largest.  It is
possible to choose its generators in such a way that stabilizers of
the larger codes are generated by appropriate subsets.  Hence the
components of the syndrome vector reflect the nested structure of the
inner code.

Again, we may not directly measure the syndromes of the $N$ copies of
the inner code.  Instead, we compute the eigenvalues using some
auxiliary quantum systems.  Then we derive syndromes for the outer
codes which will be measured.  

Details of the quantum circuits for syndrome measurement and iterative
decoding algorithms are left to further work.

\section*{Acknowledgment}
The authors would like to thank Panos Aliferis, Salman Beigi, Sergey
Bravyi, G. David Forney, Martin R{\"o}tteler, Graeme Smith, and John
Smolin for helpful discussions.

\end{document}